\documentclass[twocolumn,showpacs,amsmath,amssymb,floatfix]{revtex4}

\usepackage[dvips]{graphicx}
\usepackage[latin1]{inputenc}
\usepackage{dcolumn}
\usepackage{bm}
\begin{document}
\title{Osmotically driven flows in microchannels separated by a semipermeable membrane}
\author{
 K{\aa}re Hartvig Jensen$^{a}$,
 Jinkee Lee$^{b}$,
 Tomas Bohr$^{c}$, and
 Henrik Bruus$^{a}$}
%
%
\affiliation{%
\mbox{$^{a}$Center for Fluid Dynamics, Department of Micro- and Nanotechnology,}\\
\mbox{Technical University of Denmark, DTU Nanotech Building 345 East, DK-2800 Kongens Lyngby, Denmark}\\
\mbox{$^{b}$School of Engineering and Applied Science, Harvard University, Cambridge, Massachusetts 02138, USA}\\
\mbox{$^{c}$Center for Fluid Dynamics, Department of Physics,}\\
\mbox{Technical University of Denmark, DTU Physics Building 309, DK-2800 Kongens Lyngby, Denmark}\\
}
\date{24 October 2008}
\begin{abstract}
We perform experimental investigations of osmotically driven flows
in artificial microchannels by studying the dynamics and structure
of the front of a sugar solution traveling in $200\,\mu$m wide and
$50-200\,\mu$m deep microchannels. We find that the sugar front
travels with constant speed, and that this speed is proportional to
the concentration of the sugar solution and inversely proportional
to the depth of the channel. We propose a theoretical model, which,
in the limit of low axial flow resistance, predicts that the sugar
front indeed should travel with a constant velocity. The model also
predicts an inverse relationship between the depth of the channel
and the speed and a linear relation between the sugar concentration
and the speed. We thus find good agreement between the experimental
results and the predictions of the model. Our motivation for
studying osmotically driven flows is that they are believed to be
responsible for the translocation of sugar in plants through the
phloem sieve element cells. Also, we suggest that osmotic elements
can act as integrated pumps with no movable parts in
lab-on-a-chip systems.
\end{abstract}


\maketitle
\section{Introduction\label{sec:Introduction}}
Osmotically driven flows are believed to be responsible for the
translocation of sugar in plants, a process that takes place in the
phloem sieve element cells \cite{Taiz:2002}. These cells form a
micro-fluidic network which spans the entire length of the plant
measuring from 10 $\mu$m in diameter in small plants to 100 $\mu$m
in diameter in large trees \cite{Taiz:2002}. The mechanism driving
these flows is believed to be the osmotic pressures that build up
relative to the neighboring water-filled tissue in response to
loading and unloading of sugar into and out of the phloem cells in
different parts of the plant \cite{Taiz:2002}. This mechanism,
collectively called the pressure-flow hypothesis, is much more
efficient than diffusion, since the osmotic pressure difference
caused by a difference in sugar concentration creates a bulk flow
directed from large concentrations to small concentrations, in
accordance with the basic needs of the plant.

Experimental verification of flow rates in living plants is
difficult \cite{Knoblauch:1998}, and the experimental evidence on
artificial systems backing the pressure-flow hypothesis is scarce
and consists solely of results obtained with centimetric sized
setups \cite{Eschrich:1972,Lang:1973,Jensen:2008}. However, many
theoretical and numerical studies of the sugar translocation in
plants have used the pressure-flow hypothesis
\cite{Thompson:2003,Thompson:2003b,Holtta:2006} with good results.
To verify that these results are indeed valid, we believe that it is
of fundamental importance to conduct a systematic survey of
osmotically driven flows at the micrometer scale. Finally, osmotic
flows in microchannels can potentially act as micropumps with no
movable parts in much the same way as the osmotic pills developed
by Shire Laboratories and pioneered by F. Theeuwes
\cite{Theeuwes:1975}. Also, there is a direct analogy between
osmotically driven flows powered by concentration gradients, and
electroosmotically driven flows in electrolytes
\cite{Brask:2005,Gregersen:2007} powered by electrical potential
gradients.
\section{Experimental setup\label{sec:Experiments}}
\subsection{Chip design and fabrication\label{sec:Design}}
\begin{figure}[]
\begin{center}
\includegraphics[width=\columnwidth]{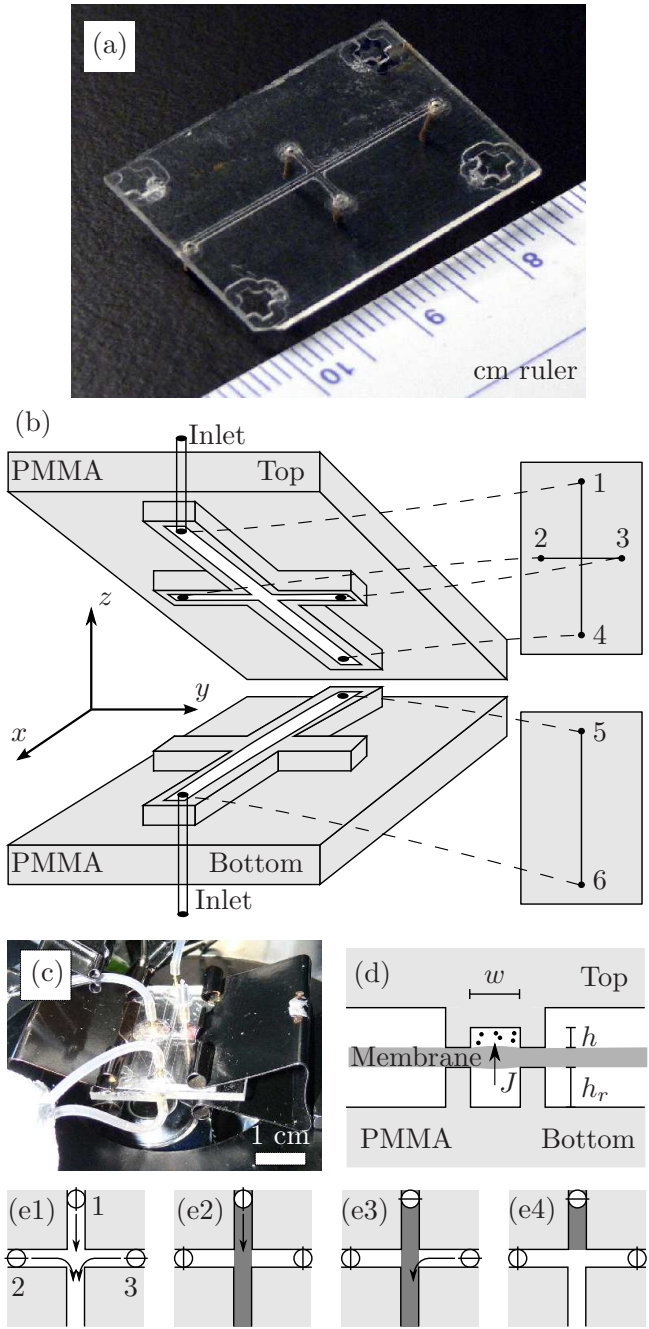}\\
\caption{(a) Picture of the top part (upside down) of the chip
showing the elevated channel and the four brass inlet tubes
(pointing down). The crosses in the four corners were used for
alignment. (b) Schematics of the two PMMA layers (gray) showing the
elevated channels (white) facing each other. All six inlet positions
(black dots) are marked, but for clarity only two brass tubes are
shown. (c) Picture of the setup assembled. (d) Schematic
cross-section closeup of the two PMMA layers (gray) clamped together
with the semipermeable membrane (dark gray) in between. The sugar in
the upper channel (black dots) and the water influx $J$ from the
lower channel (arrow) are also marked. (e1)-(e4) Valve settings
(circles) and cross-flow flushing procedure (arrows) for creating a
sharp front in the top channel between the sugar/dye solution (dark
gray) and the pure water (white). See details in the
text.\label{fig:setup}}
\end{center}
\end{figure}
To study osmotically driven flows in microchannels, we have designed
and fabricated a microfluidic system consisting of two layers of 1.5
mm thick polymethyl methacrylate (PMMA) separated by a semipermeable
membrane (Spectra/Por Biotech Cellulose Ester dialysis membrane,
MWCO 3.5 kDa, thickness $\sim 40\,\mu$m), as shown in
Fig.~\ref{fig:setup}(a)-(d). Channels of length 27~mm, width
200~$\mu$m and depth $50-200\,\mu$m were milled in the two PMMA
layers by use of a MiniMill/Pro3 milling machine \cite{Geschke:2004,
Bundgaard:2006b}. The top channel contains partly the sugar solution, and partly pure water, while the bottom channel always contains only pure water. To facilitate the production of a
steep concentration gradient by cross-flows, a 200 $\mu$m wide
cross-channel was milled in the upper PMMA layer perpendicular to
and bi-secting the main channel. Inlets were produced by drilling
$800$ $\mu$m diameter holes through the wafer and inserting brass
tubes into these.  By removing the surrounding material, the channel
walls in both the top and bottom layers acquired a height of
$100\,\mu$m and a width of $150\,\mu$m. After assembly, the two PMMA
layers were positioned such that the main channels in either layer
were facing each other. Thus, when clamping the two layers together
using two $30\,$mm paper clamps, the membrane acted as a seal,
stopping any undesired leaks from the channels as long as the
applied pressure did not exceed approximately 1 bar.

\subsection{Measurement setup and procedures \label{sec:Measurementsetupandprocedures}}
\begin{table}[t]
\centering
\begin{ruledtabular}
\caption{\label{tab:parameters} List of parameters in alphabetic order after the symbol.}
\begin{tabular}{l c l}
Parameter & Symbol & Value and/or unit  \hspace*{3.0em}\\
\hline

Initial concentration &$c_0$ & mol/L\hspace*{3.0em}\\
Diffusion constant & $D$ & m$^2$/s\hspace*{3.0em}\\
Height of channel    & $h$   & $50, 100, 200$~$\mu$m  \hspace*{3.0em}\\
Height of reservoir    & $h_r$   & $200$~$\mu$m  \hspace*{3.0em}\\
Flux across membrane & $J$ & m/s\hspace*{3.0em}\\
Length of channel   & $L$   & 27 mm \hspace*{3.0em}\\
Membrane permeability & $L_p$ & $1.8\,$ pm/(Pa$\,$s)\hspace*{3.0em}\\
Diffusion length                  & $l_D$ & m\hspace*{3.0em}\\
M\"{u}nch number    & $M$ &\hspace*{3.0em}\\
Hydrostatic pressure & $p$ & Pa \hspace*{3.0em}\\
P\'eclet number, local  &\textit{P\'e} & \hspace*{3.0em}\\
P\'eclet number, global &\textit{P\'e}$_{g}$& \hspace*{3.0em}\\
Gas constant    & $R$ & $8310\,$Pa$\,$L/(K$\,$mol)\hspace*{3.0em}\\
Position of sugar front & $s$ &\hspace*{3.0em}\\
Absolute temperature    & $T$ & K\hspace*{3.0em}\\
Time                  & $t$ & s\hspace*{3.0em}\\
$x$-velocity of sugar front  & $U$   & m/s \hspace*{3.0em}\\
Mean $x$-velocity of sugar front  & $u$   &  m/s\hspace*{3.0em}\\
Volume behind sugar front &$V$& m$^3$ \hspace*{3.0em}\\
Width of channel    & $w$   & 200 $\mu$m \hspace*{3.0em}\\
Width of sugar front & $w_f$ & m\hspace*{3.0em}\\
Cartesian coordinates & $x,y,z$& m\hspace*{3.0em}\\
Osmotic coefficients: &  &\hspace*{3.0em}\\
\hspace*{1.5em}Dextran ($T=293\,$K)  & $\alpha$&41, see Ref. \cite{Jensen:2008} \hspace*{3.0em}\\
\hspace*{1.5em}Sucrose ($T=293\,$K)  & $\alpha$&1, see Ref. \cite{Michel:2003} \hspace*{3.0em}\\
Viscosity   & $\eta$ & Pa$\,$s\hspace*{3.0em}\\
Position of sugar front   & $\lambda$   &  m\hspace*{3.0em}\\
Position of initial sugar front  & $\lambda_0$   & 13.5 mm \hspace*{3.0em}\\
Osmotic pressure & $\Pi$ & Pa\hspace*{3.0em}\\
\end{tabular}
\end{ruledtabular}
\end{table}
In our setup, the osmotic pressure pushes water from the lower
channel, through the membrane, and into the sugar-rich part of the
upper channel. This displaces the solution along the upper channel
thus generating a flow there. To measure this flow inside the upper
channel, particle and dye tracking were used. In both cases inlets
1, 2, 3 and 5 (see Fig.~\ref{fig:setup}) were connected via silicone
tubing (inner diameter 0.5 mm) to disposable syringes. Syringes~2, 3
and~5 was filled with demineralised water and syringe~1 was filled
with a solution of sugar (sucrose or dextran (mol. weight: 17.5~kDa,
Sigma-Aldrigde, type D4624)) and 5~$\%$ volume red dye (Flachsmann
Scandinavia, Red Fruit Dye, type 123000) in the dye tracking
experiments and 0.05~$\%$ volume sulfate modified  $1$~$\mu$m
polystyrene beads (Sigma Aldrigde, L9650-1ML) in the particle
tracking experiments. Inlets 4 and 6 were connected to the same
water bath to minimize the hydrostatic pressure difference between
the two sides of the membrane.

\begin{figure}[]
\begin{center}
\includegraphics[width=80mm]{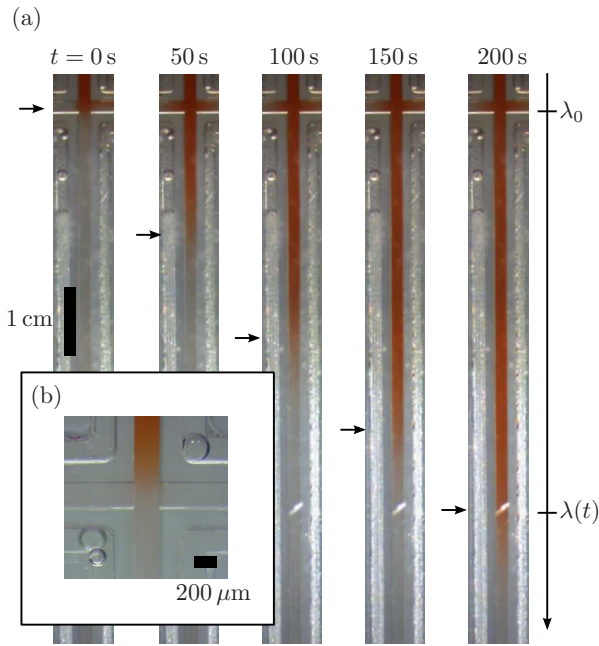}
\caption{(a): Images showing the sugar front moving in the
$200\,\mu$m$\times200\,\mu$m channel. The time between each image is 50~s. The arrows indicate the position of the sugar front as it moves down along the channel. (b): Closeup of the
cross junction just after a sharp sugar/water interface has been created.\label{fig:dye_tracking_data_images}}
\end{center}
\end{figure}

When conducting both dye tracking and particle tracking experiments,
the initialization procedure shown in Fig.~\ref{fig:setup}(e1)-(e4)
was used: First (e1), inlet valves 1, 2 and 3 were opened and all
channels were flushed thoroughly with pure water (white) to remove
any air bubbles and other impurities. Second (e2), after closing
inlets 2 and 3 a sugar solution (dark gray) was injected through
inlet 1 filling the main channel in the upper layer. Third (e3),
inlet 1 was closed and water was carefully pumped through inlet 2 to
produce a sharp concentration front at the cross, as shown in
Fig.~\ref{fig:setup}(e4) and~\ref{fig:dye_tracking_data_images}(b).

\subsubsection{Sugar front motion recorded by dye tracking}
The motion of the sugar front in the upper channel was recorded by
taking pictures of the channel in 10~s intervals using a Leica MZ 16
microscope. This yielded images as those displayed in
Fig.~\ref{fig:dye_tracking_data_images}(a), clearly showing a front
(marked by arrows) of the sugar/dye solution moving along the
channel. To obtain the position $\lambda (t)$ of the sugar front as
a function of time $t$, the distance from the initial front position
$\lambda_0$ to the current position $\lambda (t)$ was measured using
ImageJ software. The position of the sugar front was taken to be at
the end of the highly saturated dark region. In this way, the
position of the front could be measured at each time step with an
accuracy of $\pm $200~$\mu$m. As verified in earlier works \cite{Eschrich:1972,Jensen:2008}, we assumed that the sugar and dye
traveled together, which is reasonable since the  P\'eclet number is
$P\acute e \sim 10$ (see Section~\ref{sec:Theory}). We only
applied the dye tracking method on the $200$~$\mu$m deep channel,
since the $100$~$\mu$m and $50$~$\mu$m deep channels were too
shallow for sufficient scattering of red light by the solution to
get a clear view of the front.

\subsubsection{Sugar front motion recorded by use of particle tracking}
\begin{figure}[]
\begin{center}
\includegraphics[width=80mm]{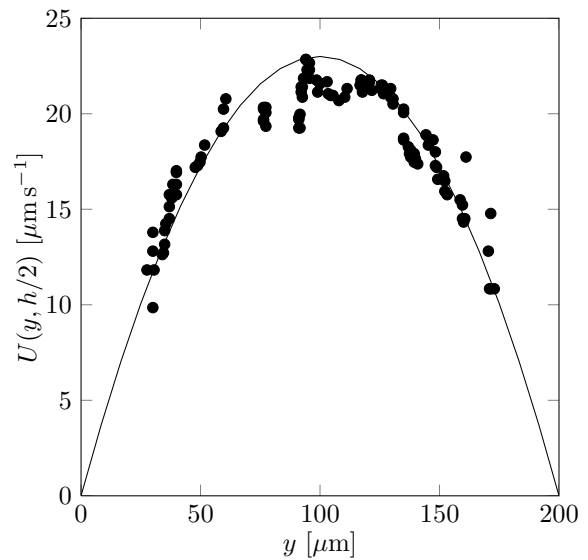}
\caption{Velocity profile $U(y,h/2)$ obtained in the center plane across the $200\,\mu$m$\times 200\,\mu$m channel. The solid black line is a fit to Eq.~(\ref{eq:bruus2}). \label{fig:particle_tracking_velocity_profile}}
\end{center}
\end{figure}
The flow velocity inside the upper channel was recorded by tracking
the motion of $1$~$\mu$m beads in the water $3$~mm ahead of the
initial sugar front position. Images were recorded every
$200-1000$~ms using a Unibrain Fire-i400 1394 digital camera
attached to a Nikon Diaphot microscope with the focal plane at
$h/2$. From the images we extracted velocity profiles such as the
one shown in Fig.~\ref{fig:particle_tracking_velocity_profile}. At
the point of observation well ahead of the front, the flow behaves
as if it were pressure driven. In that case the laminar flow profile
$U$ in the rectangular straight top channel of height $h$, width
$w$, and length $L$ is given by the expression \cite{Bruus:2008}
\begin{equation}
U(y,z)=
\frac{4h^2\Delta p}{\pi^3\eta L}\sum_{n, \text{odd}}^\infty\frac{1}{n^3}\left[1-\frac{\cosh \left(n\pi \frac yh\right)}{\cosh \left(n\pi \frac w{2h}\right)}\right]\sin\left(n\pi\frac zh\right).
\label{eq:bruus1}
\end{equation}
At the center of the channel, $z=h/2$,  this simplifies to
\begin{equation}
U(y,h/2)=
\frac{4h^2\Delta p}{\pi^3\eta L}\sum_{n, \text{odd}}^\infty\frac{(-1)^n}{n^3}\left[1-\frac{\cosh \left(n\pi \frac yh\right)}{\cosh \left(n\pi \frac w{2h}\right)}\right].
\label{eq:bruus2}
\end{equation}
To determine the pre-factor, we fit Eq.~(\ref{eq:bruus2}) to our
data points obtained by particle tracking. The average velocity $u$
inside the channel is then
\begin{equation}
u=\frac {h^2\Delta p}{12\eta L}\left[1-\sum_{n,\text{odd}}\frac{1}{n^5}\frac{192}{\pi^5}\frac hw\tanh \left(n\pi\frac{w}{2h}\right)\right].
\end{equation}

\section{Experimental results}

\subsection{Dye tracking}
\begin{figure}[]
\begin{center}
\includegraphics[width=80mm]{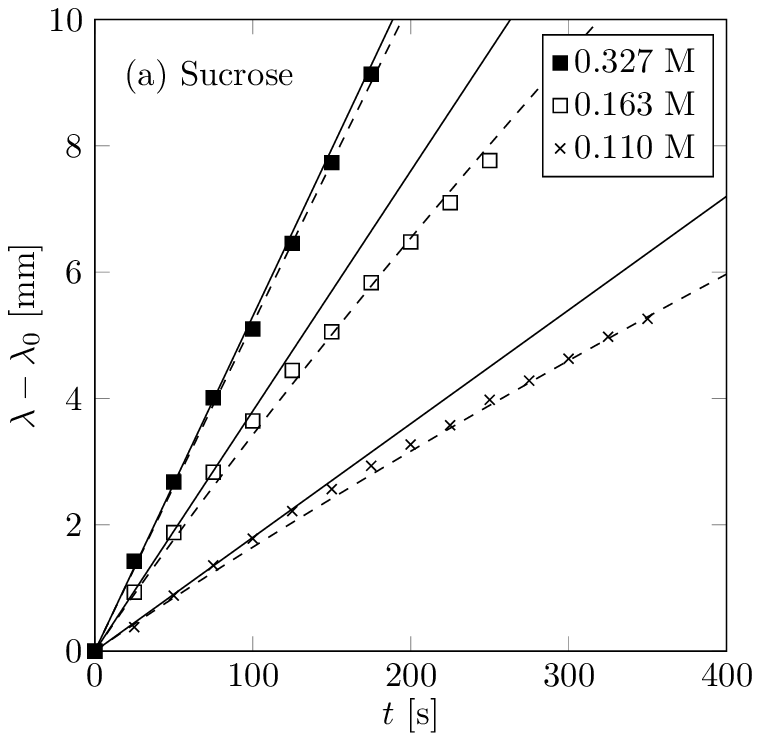}\\
\includegraphics[width=80mm]{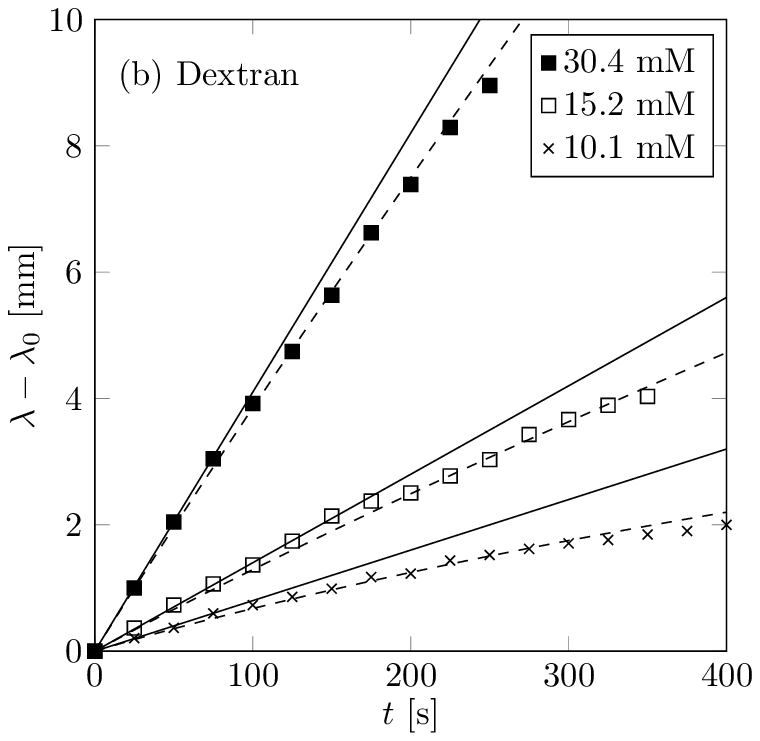}\\
\includegraphics[width=80mm]{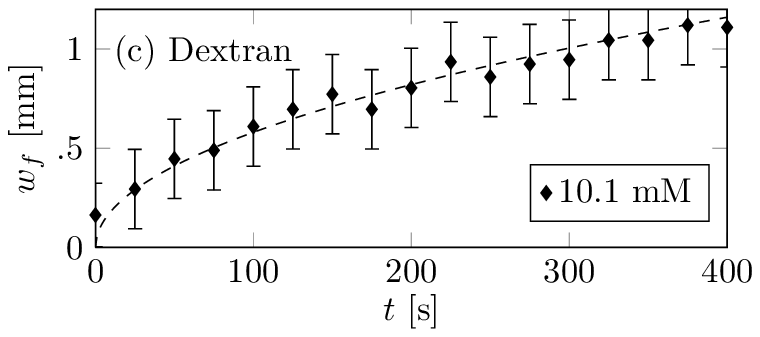}\\
\caption{Measured position $\lambda$ of the sugar front as a
function of time $t$ in the $200\,\mu$m$\times200\,\mu$m channel for
various concentrations of (a) sucrose and (b) dextran. The solid
black lines are linear fits for $0\;$s$\,<\,t\,<100\,$s. The dashed
lines are fits to Eq.~(\ref{eq:corrsolution02}). (c) The width $w_f$
of the sugar front as a function of time for the 10.1 mM dextran
experiment. The dashed black line is a fit to
$\left(2Dt\right)^{1/2}$ with $D=1.7\times
10^{-9}\,$m$^2\,$s$^{-1}$.\label{fig:dye_tracking_raw_data}}
\end{center}
\end{figure}
Figure \ref{fig:dye_tracking_raw_data} shows the position of the
sugar front in the $200$~$\mu$m deep channel as a function of time obtained by dye tracking. The data sets correspond to different
concentrations of sucrose and dextran as indicated in the legends.
Initially, the sugar front moves with constant speed, but then it
gradually decreases, more so for low than high concentrations. The
solid black lines are linear fits for the first 100 s giving the
initial velocity of the front. In
Fig.~\ref{fig:dye_tracking_raw_data}(c) the width of the sugar front
for the 10.1 mM dextran experiment is shown along with a fit to
$\left(2Dt\right)^{1/2}$ showing that the sugar front broadens by
molecular diffusion.

\subsection{Particle tracking}
\begin{figure}[]
\begin{center}
\includegraphics[width=80mm]{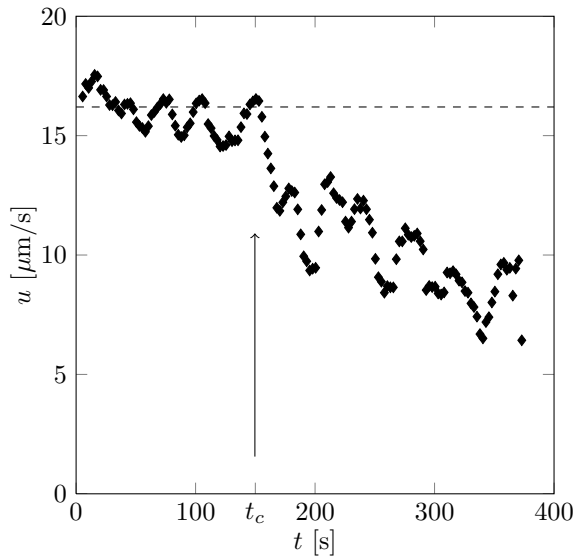}\\
\caption{The average flow velocity $u$ in the 200 $\mu$m deep channel
as a function of time $t$ measured 3 mm ahead of the initial front
position. At $t_c\simeq 150\,$s the sugar front begins to reach the
observation point, and the velocity decreases rapidly. For $t>t_c$,
the velocity was not determined
accurately.\label{fig:particle_tracking_raw_data}}
\end{center}
\end{figure}
\begin{figure}[]
\begin{center}
\includegraphics[width=80mm]{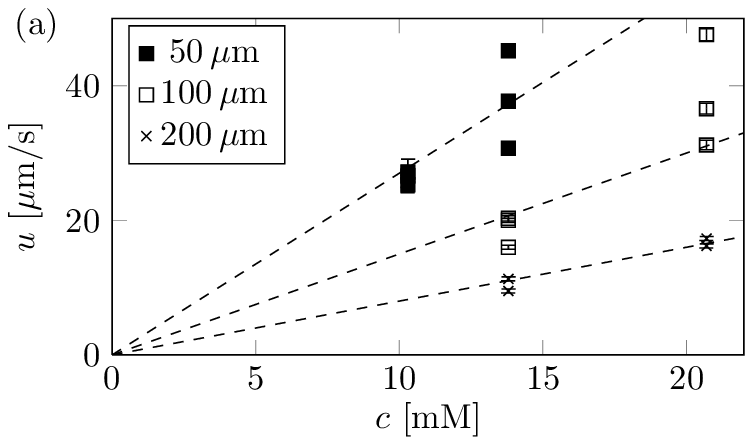}\\
\includegraphics[width=80mm]{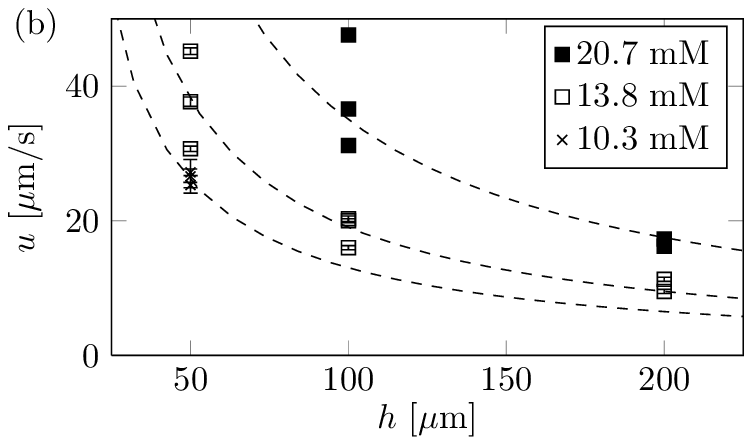}\\
\caption{Front velocity $u$ obtained by particle tracking. (a) The
velocity $u$ plotted against dextran concentration $c_0$. The dashed
lines are fits to $c$ provided as guides to the eye. (b) The
velocity $u$ plotted against channel depth $h$. The dashed lines are
fits to $1/h$ provided as guides to the
eye.\label{fig:particle_tracking_data}}
\end{center}
\end{figure}
Figure \ref{fig:particle_tracking_raw_data} shows the velocity as a
function of time obtained by particle tracking in a
$200\,\mu$m$\times200\,\mu$m channel. For the first $150$~s the
velocity is approximately constant after which it starts decreasing
as the sugar front passes the point of observation. We interpret the
mean value of the initial plateau of the velocity graph as the speed
of the sugar front. Figs.~\ref{fig:particle_tracking_data}(a)
and~(b) shows the velocity of the sugar front as a function of dextran
concentration and of channel depth obtained in this way.

\section{Theoretical analysis}
\label{sec:Theory} When modeling the flow inside the channel, we use
an approach similar to that of Eschrich \emph{et al.}
\cite{Eschrich:1972}. They introduced a 1D model with no axial flow
resistance and zero diffusivity in a setting very similar to ours.
To formalize this, we consider the two most important
non-dimensional numbers in the experiments: the M\"{u}nch number $M$ \cite{Jensen:2008} and the P\'eclet number $P\acute e$ \cite{Bruus:2008}. These numbers characterize the ratio of axial to
membrane flow resistance and axially convective to diffusive fluxes
respectively. In our experiments
\begin{equation}
M=\frac{\eta L^2L_p}{h^3}\sim  10^{-6},
\end{equation}
and
\begin{equation}
P\acute{e}=\frac{w_fu}{D}\sim 10.
\end{equation}
Here $\eta$ is the viscosity (typically $1.5$~mPa$\,$s), $w_f$ is
the front width (typically $500$ $\mu$m), and $D$ the molecular diffusivity of
sugar (typically $10^{-10}$ m$^2$s$^{-1}$ for sucrose and the dye
and $10^{-11}$ m$^2$s$^{-1}$ for dextran)

\subsection{Equation of motion}
Since $M\ll 1$ and $Pe\gg1$, we shall neglect the axial flow
resistance and the diffusion of the sugar in our analysis. In this
way, let $\lambda(t)$ denote the position of the sugar/dye front in
the upper channel, and let $V$ denote the volume behind the front.
The flux $J$ of water across the membrane from the lower to the
upper channel, see Fig.~\ref{fig:setup}(d), is given by
\begin{equation}
J=L_p\left(\Delta p + \Delta \Pi\right)\simeq L_p\alpha RTc,
\end{equation}
where $L_p$ is the membrane permeability, $\Delta p$ the hydrostatic and $\Delta \Pi$ the osmotic pressure difference across the membrane. In our experiments $\Delta p=0$, and from the van 't Hoff relation follows $\Delta \Pi\simeq \alpha RTc$, where $\alpha$ is the osmotic coefficient, $R$ is the gas constant, $T$ is the absolute temperature, and $c$ is the concentration of sugar molecules.  Since the concentration is independent of $x$ behind the front and zero ahead of it, $J$ is also independent of $x$. By the conservation of sugar this allows us to write the concentration as
\begin{equation}
c(x,t)=
\begin{cases}
c_0\frac{\lambda_0}{\lambda(t)} &
x  \leq \lambda(t),\\
0 & x \geq  \lambda(t).\\
\end{cases}
\label{eq:conc1}
\end{equation}
Moreover, the rate of change of the expanding volume $V$ behind the front can be related to $J$ as
\begin{eqnarray}
\frac{\mathrm d V}{\mathrm d t}   &=& w\int_0^L J(x)dx\label{eq:dV1}\nonumber\\
                                &=& wL_p\alpha RTc_0\frac{\lambda_0}{\lambda(t)}\int_0^{\lambda (t)}\mathrm dx\nonumber\\
                                &=& w\lambda_0L_p\alpha RTc_0\label{eq:volume1}.
\end{eqnarray}
However, we also have that
\begin{equation}
\frac{\mathrm dV}{\mathrm dt}=hw\frac{\mathrm d \lambda(t)}{\mathrm d t}\label{eq:dV2},
\end{equation}
which implies together with Eq.~(\ref{eq:volume1}) that
\begin{equation}
\lambda(t)=\lambda_0+\frac{\lambda_0}{h}L_p\alpha RTc_0t=\lambda_0+ut,
\end{equation}
where the velocity $u$ of the front is given by
\begin{equation}
u=\frac{\lambda_0}{h}L_p\alpha RTc_0.
\label{eq:frontvel}
\end{equation}
\subsection{Corrections to the equation of motion}
In the previous section, we considered the motion of a sharp sugar
front, as given by the stepwise concentration profile in
Eq.~(\ref{eq:conc1}), and found that this moved with constant
velocity. However, as can be seen in
Fig.~\ref{fig:dye_tracking_raw_data} the front velocity gradually
decreases. To explain this, we observe that in
Fig.~\ref{fig:dye_tracking_data_images}(a) there exist a region of
growing size separating the sugar-filled region from the region of
pure water. This intermediate region is caused by diffusion and
hence denoted the diffusion region. The end point of the sugar
region is denoted $\lambda$, while the width of the diffusion
region, which is growing in time, is denoted $l_D$. Consequently,
the concentration profile can be approximated by the following
simple three-region model,
\begin{equation}
c(x,t)=
\begin{cases}
c_t, & 0\leq x\leq \lambda,\\[2mm]
c_t\left(1-\frac{x-\lambda}{l_D}\right),&\lambda\leq x\leq \lambda+l_D,\\[2mm]
0, & \lambda+l_D\leq x\leq L.\\
\end{cases}
\label{eq:conc2}
\end{equation}
Here $l_D=(2Dt)^{1/2}$ and $c_t=c_0\frac{\lambda_0}{\lambda+l_D/2}$
from conservation of sugar. Using Eqs.~(\ref{eq:dV1})
and~(\ref{eq:dV2}) the time derivative of $\lambda$ becomes
\begin{equation}
\frac{\mathrm d\lambda}{\mathrm dt}=\frac{L_p \alpha RTc_0\lambda_0}{h}\frac{\lambda}{\lambda+l_D}.
\end{equation}
Rescaling using $\lambda=s\lambda_0$ and
$t=\tau\frac{\lambda_0}{u}$, we get that
\begin{equation}
\frac{\mathrm ds}{\mathrm
d\tau}=\frac{s}{s+\left(\frac{\tau}{P\acute{e}_g}\right)^{1/2}},
\label{eq:corrsolution01}
\end{equation}
where we have introduced the global P\'eclet number
\begin{equation}
P\acute{e}_g=\frac{2\lambda_0^2L_p\alpha R T c_0}{Dh}=\frac{2\lambda_0}{D}u.
\end{equation}
Given the experimental conditions, $P\acute{e}_g$ is typically of the order
$10^1-10^2$. Thus, for $\left(\frac{\tau}{P\acute{e}_g}\right)^{1/2}\ll
1$, Eq.~(\ref{eq:corrsolution01}) can be solved by an expansion,
\begin{equation}
s=s_0+\tau\left(1-\frac{2}{3s_0}\left(\frac{\tau}{P\acute{e}_g}\right)^{1/2}+
\mathcal{O}\left[\left(\frac{\tau}{P\acute{e}_g}\right)\right]\right).
\label{eq:corrsolution02}
\end{equation}
The dashed lines in Figs.~\ref{fig:dye_tracking_raw_data}(a) and~(b)
are fits to Eq.~(\ref{eq:corrsolution02}), with values of $D$
between $2\times 10^{-7}\,$m$^2\,$s$^{-1}$ and
$4\times10^{-9}\,$m$^2\,$s$^{-1}$, showing good qualitative
agreement between theory and experiment. However, these values of
$D$ are 1 to 100 times larger than that obtained  in
Fig.~\ref{fig:dye_tracking_raw_data}(c) ($1.7\times
10^{-9}\,$m$^2\,$s$^{-1}$) indicating a quantitative discrepancy
between the experimental data and our model. We suspect that this is
due to some accelerated diffusion mechanism occurring at the front,
such as Taylor dispersion \cite{Taylor:1953, Lee:2008}. To resolve
this issue, experiments of higher accuracy are required as well as
direct tracking of the sugar without using a dye.

\section{Discussion} \label{sec:Discussion}
\subsection{Comparison of theory and experiment}
\begin{figure}[]
\begin{center}
\includegraphics[width=\columnwidth]{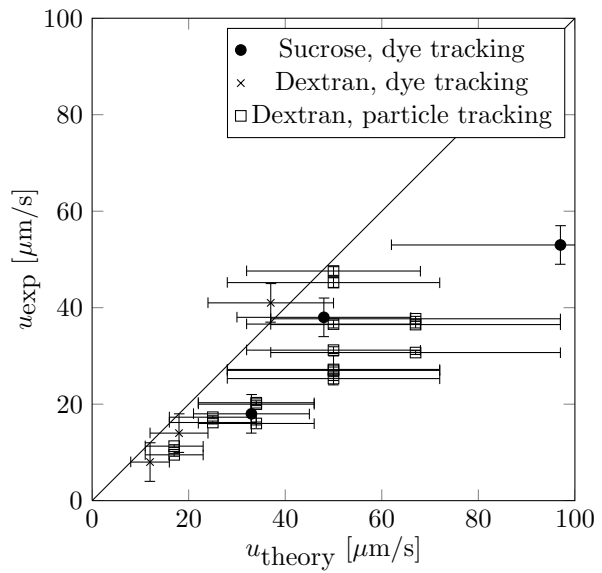}
\caption{The experimental values of $u$ plotted against the results from Eq.~(\ref{eq:frontvel}).\label{fig:velocitycompare}}
\end{center}
\end{figure}
To compare the experimental data with theory, we have in
Fig.~\ref{fig:velocitycompare} plotted the empirically obtained
velocities $u_{\rm exp}$ against those predicted by
Eq.~(\ref{eq:frontvel}). For nearly all the dextran and sucrose
experiments we see a good agreement between experiment and theory,
although Eq.~(\ref{eq:frontvel}) systematically overestimates the
expected velocities.

We interpret the quantitative disagreement as an indication of a
decreasing sugar concentration in the top channel due to diffusion
of sugar into the membrane as well as the presence of a
low-concentration boundary layer near the membrane, a so-called
unstirred layer \cite{Pedley:1983}.

\subsection{Osmotic pumps in lab-on-a-chip systems}
Depending on the specific application, flows in lab-on-a-chip
systems are conventionally driven by either syringe pumps or by
using more advanced techniques such as electronically controlled
pressure devices, electro-osmotic pumps \cite{Ajdari:2000},
evaporation pumps \cite{Noblin:2008}, or capillary pumps
\cite{Boudait:2005}. Most of these techniques involves the
integration of either movable parts or complicated electronics into
the lab-on-a-chip device. As an application of our results, we
suggest the use of osmotic pumps in lab-on-a-chip systems. This
could be done by integrating in the device a region where the
channel is in contact through a membrane with a large reservoir
containing an osmotically active agent. By using a sufficiently
large reservoir, say $1\,$cm$^{3}$, and a $100\,\mu$m$\times
100\,\mu$m channel with a flow rate of $100\,\mu$m/s it would take
more than $10$ days to reduce the reservoir concentration by $50$\%
and thus decreasing the pumping rate by 50\%. We emphasize that such
osmotic pumping would be completely steady, even at very low flow
rates.
\section{Conclusions\label{sec:Conclusions}}
We have studied osmotically driven, transient flows in 200 $\mu$m
wide and $50-200$ $\mu$m deep microchannels separated by a
semipermeable membrane. These flows are generated by the influx of
water from the lower channel, through the membrane, into the large
sugar concentration placed in one end of the top channel. We have
observed that the sugar front in the top channel travels with
constant speed, and that this speed is proportional to the
concentration of the sugar solution and inversely proportional to
the depth of the channel. We propose a theoretical model, which, in
the limit of low axial flow resistance, predicts that the sugar
front should travel with a constant velocity. The model also
predicts an inverse relationship between the depth of the channel
and the speed and a linear relation between the sugar concentration
and the speed. We compare theory and experiment with good agreement,
although the detailed mechanism behind the deceleration of the flow
is still unknown. Finally, we suggest that osmotic elements can
potentially act as pumps with no movable parts in lab-on-a-chip
systems.

\section{Acknowledgements}
It is a pleasure to thank Emmanuelle Rio, Christophe Clanet, Frederik Bundgaard, Jan
Kafka and Oliver Geschke for assistance and advice on chip design and
manufacturing. We also thank Alexander Schulz, Michele Holbrook,
Maciej Zwieniecki and Howard Stone for many useful discussions of
the biological and phycial aspects of osmotically driven flows. This
work was supported by the Danish National Research Foundation, Grant
No. 74 and by the Materials Research Science and Engineering Center
at Harvard University.
\bibliography{KHJ_microosmosis}
\end{document}